\documentclass[twocolumn]{aastex631}

\begin{document}

\title{Sudden extreme obscuration of a Sun-like main-sequence star: evolution of the circumstellar dust around ASASSN-21qj}

\correspondingauthor{Jonathan P. Marshall}
\email{jmarshall@asiaa.sinica.edu.tw}

\author[0000-0001-6208-1801]{Jonathan P. Marshall}
\affiliation{Academia Sinica Institute of Astronomy and Astrophysics, 11F of AS/NTU Astronomy-Mathematics Building,\\No.1, Sect. 4, Roosevelt Rd, Taipei 10617, Taiwan}
\affiliation{University of Southern Queensland, Centre for Astrophysics, USQ Toowoomba, West Street, QLD 4350, Australia}

\author[0000-0002-2314-7289]{Steve Ertel}
\affiliation{Department of Astronomy and Steward Observatory, University of Arizona, 933 N. Cherry Avenue, Tucson, AZ 85721-0065, USA}
\affiliation{Large Binocular Telescope Observatory, University of Arizona, 933 N. Cherry Avenue, Tucson, AZ 85721-0065, USA}

\author[0000-0003-2743-8240]{Francisca Kemper}
\affiliation{Institut de Ci\`encies de l'Espai (ICE, CSIC), Can Magrans, s/n, E-08193 Cerdanyola del Vall\`es, Barcelona, Spain}
\affiliation{ICREA, Pg. Llu\'{\i}s Companys 23, E-08010 Barcelona, Spain}
\affiliation{Institut d'Estudis Espacials de Catalunya (IEEC), E-08034 Barcelona, Spain}

\author[0000-0002-8949-5200]{Carlos del Burgo}
\affiliation{Instituto Nacional de Astrof\'isica, \'Optica y Electr\'onica, Luis Enrique Erro \#1, CP 72840, Tonantzintla, Puebla, Mexico} 

\author[0000-0002-6717-1977]{Gilles P. P. L. Otten}
\affiliation{Academia Sinica Institute of Astronomy and Astrophysics, 11F of AS/NTU Astronomy-Mathematics Building,\\No.1, Sect. 4, Roosevelt Rd, Taipei 10617, Taiwan}

\author[0000-0002-1161-3756]{Peter Scicluna}
\affiliation{European Southern Observatory, Alonso de Cordova 3107, Santiago RM, Chile}

\author[0000-0002-8163-8852]{Sascha T. Zeegers}
\affiliation{European Space Agency, ESTEC/SRE-SA, Keplerlaan 1, 2201 AZ, Noordwijk, The Netherlands}

\author[0000-0003-3133-3580]{\'Alvaro Ribas}
\affiliation{Institute of Astronomy, University of Cambridge, Madingley Road, Cambridge CB3 0HA, UK}

\author[0000-0002-5908-9543]{Oscar Morata}
\affiliation{Institut de Ci\`encies de l'Espai (ICE, CSIC), Can Magrans, s/n, E-08193 Cerdanyola del Vall\`es, Barcelona, Spain}



\begin{abstract}
ASASSN-21qj is a distant Sun-like star that recently began an episode of deep dimming events after no prior recorded variability. Here we examine archival and newly obtained optical and near-infrared data of this star. The deep aperiodic dimming and absence of previous infrared excess are reminiscent of KIC 8462852 (``Boyajian's Star''). The observed occultations are consistent with a circumstellar cloud of sub-micron-sized dust grains composed of amorphous pyroxene, with a minimum mass of $1.50~\pm~0.04\times10^{-9}~M_{\oplus}$ derived from the deepest occultations, and a minimum grain size of $0.29^{+0.01}_{-0.18}~\mu$m assuming a power law size distribution. We further identify the first evidence of near-infrared excess in this system from NEOWISE 3.4 and 4.6~$\mu$m observations. The excess emission implies a total circumstellar dust mass of around $10^{-6} M_{\oplus}$, comparable to the extreme, variable discs associated with terrestrial planet formation around young stars. The quasiperiodic recurrence of deep dips and the inferred dust temperature (ranging from 1800 to 700~K across the span of observations) independently point to an orbital distance of $\simeq$0.2~au for the dust, supporting the occulting material and excess emission being causally linked. The origin of this extended, opaque cloud is surmised to be the breakup of one or more exocometary bodies. 
\end{abstract}



\section{Introduction}
\label{sec:introduction}

Stars of all ages and masses exhibit variability; the source of their variability can be either intrinsic to the star, for example pulsations or surface features, or extrinsic and arising from circumstellar material, for example occultation by a disc. Pre-main sequence stars form surrounded by massive and opaque gas- and dust-rich protoplanetary discs within which planetesimals, and planets, form \citep{2001Haisch,2007Hernandez,2015Ribas}. These systems evolve over time into lower mass, more tenuous debris discs wherein the circumstellar material is (predominantly) derived from mutual collisions within belts of planetesimals \citep{2015Wyatt,2018Hughes}. Protoplanetary discs may be clumpy, warped, or inclined in such a way that star light reaches us attenuated by different amounts over time. Dimming from such systems can be up to several magnitudes and last several months \citep[e.g.][]{2017aRodriguez,2017bRodriguez}. Such disc-induced variable systems are relatively common, comprising $\simeq 30-40\%$ of young systems, and are collectively known as ``dippers'' \citep{2010Alencar,2011MoralesCalderon,2016Andsell,2020Andsell,2022Capistrant}.

Debris discs are composed of dusty and icy bodies, from kilometer-size planetesimals to micron-size dust, created in belts around their host stars \citep{2018Hughes}. A handful of debris disc systems exhibit evidence for the existence of star-grazing planetesimal populations, exocomets, flung onto eccentric orbits far from their birthplace \citep{2020Strom}. The most extreme examples attributed to possible exocometary behaviour are ``little dippers'', stars exhibiting episodic, asymmetric transits. The dimming of ``little dippers'' is most commonly associated with stellar activity or a mis-aligned inner disc, but some of the dimming events of these systems could be attributed to exocometary phenomena making them a useful point of comparison \citep{2016Scaringi,2019Andsell,2022Pavlenko}.

The main sequence star with the most extreme variability associated with exocometary phenomena is KIC 8462852 (``Boyajian's star''), which showed significant dimming at irregular intervals over a $\sim$ 1\,500 day period \citep{2016Boyajian}. Short events blocking more than 20 \% of the starlight have been observed, as well as prolonged asymmetric obscurations of a few \%, with durations of 10 to 20 days \citep{2016Montet,2017Meng}. Multi-wavelength photometric observations of KIC 8462852, exhibiting a wavelength dependent transit depth, are consistent with an optically thin clump of sub-micron-sized dust grains \citep{2018Boyajian,2018Deeg,2019Hitchcock}. Frequency analysis of the dimming is inconclusive on the periodicity of KIC 8462852, suggesting a range up to several years \citep{2017Kiefer,2018Ballesteros,2018Bourne}. The recent recurrence of activity favours a periodicity of $\sim$ 4.4 years, suggesting the occulting material lies in the inner regions of that system \citep{2018Castelaz}. 

A few dozen little dippers have been identified in long term photometric monitoring programs, exhibiting similar aperiodic occultations at the few percent level which unfold over several days \citep{2019Schmidt,2019Andsell,2019Kennedy,2019aRappaport,2019bRappaport}. A population of these peculiar systems is now emerging \citep{2022Schmidt}. Several possible causes for their unusual optical light curves have been explored \citep{2016Wright}, with a stream of exocomets orbiting the host star being the most plausible explanation \citep{2016Boyajian}. This idea has received support from calculations estimating the disruption of a Ceres-like dwarf planet into smaller fragments, or the start of a late heavy bombardment type event \citep{2016Bodman}. The presence of deep transits associated with circumstellar dust could be associated with an infrared excess. However, hitherto no infrared excesses have been detected amongst the known sample of little dippers, likely due to the low dust masses associated with the transits ($\sim 10^{-4}~M_{\rm Earth}$) and large distances to these stars ($> 100~$pc) \citep{2018Wyatt}.

A drastic drop in the brightness of Gaia DR3 5539970601632026752 \citep{2021Gaia}, hereafter ASASSN-21qj, was recently reported \citep{2021ATel14879....1R,2022ATel15531....1R}, recording an unprecedented reduction in the star from $g^{\prime}$ 13.8 mag to $>$16 mag (at the limit of ASAS-SN sensitivity) over several weeks, after previously exhibiting no strong variability in several years of monitoring by the All-Sky Automated Survey for SuperNovae \citep[ASAS-SN,][]{2014Shappee,2017Kochanek}. Following an extended period of deep dimming, the star returned with slightly lower brightness and redder colour than in its pre-dimming phase. The amplitude of the variability is larger in the visible than in the near-infrared, suggestive of occultation by a dust cloud \citep{2022ATel15531....1R}. Here we present previously unpublished optical and near-infrared observations tracing the evolution of this event after its discovery, seeking to characterize both the spatial distribution of circumstellar material in the system, and identify the size, composition, and mass of the constituent dust grains responsible for the dimming.

The remainder of the paper is laid out as follows. In Section \ref{sec:obs} we present the monitoring observations used to trace the evolution of this event and the ancilliary data used to interpret the nature of the dust. We derive the fundamental stellar parameters of ASASSN-21qj and describe the lightcurve modelling to measure the magnitude of the extinction using $g^{\prime}$ photometry, calculate the grain size and composition of the occulting dust from the observed reddening, and thereafter infer the mass of dust responsible, along with a review of infrared emission to spot an excess in Section \ref{sec:res}. We discuss the significance of both detections in relation to the ensemble of previously identified dimming stars and give our conclusions in Section \ref{sec:con}.

\section{Observations} 
\label{sec:obs}

We monitored ASASSN-21qj with time series photometry from 11~September~2021 to 20~June~2022 (PID: DDT2021B-003, PI: S. Ertel, PID: NSF2022A-010, PI: J. P. Marshall) using the Las Cumbres Observatory's \citep[LCOGT,][]{2013Brown} network of 1-m telescopes with the SINISTRO imaging instrument. The stellar brightness was recorded in four filter bands: SDSS $g^{\prime}$, $r^{\prime}$, $i^{\prime}$ and Pan-STARRS $z_{s}$. Integration times were 18s in $g^{\prime}$, 20s in $r^{\prime}$, 30~s in $i^{\prime}$, and 60~s in $z_{s}$. This provides sufficient signal-to-noise for a 5-$\sigma$ detection if the star dimmed to 20th magnitude (from an unocculted $g^{\prime}$ magnitude of 13.8). The star was observed every other night in $g^{\prime}$, $r^{\prime}$, and $i^{\prime}$, with additional $z_{s}$ triggered upon request up to 31~January~2022, and thereafter every other night in $g^{\prime}$ with additional measurements in $r^{\prime}$, $i^{\prime}$, and $z_{s}$ triggered upon request. Photometric measurements were taken from the pipeline produced BANZAI catalogue \citep{2018McCully,2021Xu}. If the catalogue (or target) was missing from the observatory provided data, the stellar magnitude was determined using aperture photometry. 

Aperture photometry was performed using the Python-based {\sc Photutils} package \citep{2022Bradley} using a 5$\arcsec$ radius aperture and a background annulus between 10$\arcsec$ to 15$\arcsec$. There is a companion star at a separation of $4\arcsec$, whose contribution to the emission within the aperture was subtracted using a scaled PSF. Magnitudes calculated using aperture photometry were cross-checked for consistency with BANZAI-derived values. The LCOGT measurements are presented in Figure \ref{fig:timeseries}, along with preceding publicly available ASAS-SN $V$ and $g^{\prime}$, and ATLAS `cyan' (653~nm) measurements\footnote{The LCOGT and ATLAS measurements presented in this paper can be downloaded \href{https://github.com/jontymarshall/Extreme_Occultations_of_ASAS_SN-21qj}{here}. ASAS-SN measurements can be obtained from the transients webpage \href{https://www.astronomy.ohio-state.edu/asassn/transients.html}{here}.} \citep{2014Shappee} and contemporaneous near-infrared NEOWISE W1 and W2 measurements \citep{2014Mainzer,neowise}. 

\section{Analysis and Results}
\label{sec:res}

\subsection{Fundamental stellar parameters}

In order to derive the fundamental stellar parameters of ASASSN-21qj, we applied the Bayesian inference code of \citet{2016delBurgo,2018delBurgo} on the PARSEC v1.2S library of stellar evolution models \citep{2012Bressan}. This combination is an informed choice given the good statistical agreement of the dynamical masses of detached eclipsing binaries with their respective inferences, especially for main-sequence stars, where predicted masses are, on average, 4~\% accurate \citep{2018delBurgo}.

We follow the method of \citet{2016delBurgo}, taking the absolute $G$ magnitude, $M_{G}$, the colour $G_{BP}-G_{RP}$ from \textit{Gaia} DR3 \citep{2021Gaia}, together with an abundance ratio of iron to hydrogen [Fe/H]=~$-0.20~\pm~0.20$ dex. 
We obtained $M_{G}$ using the geometric distance for ASASSN-21qj of 556$^{+3}_{-4}$ pc \citep{2021BailerJones}. 

The given metallicity is favoured by our overall prediction and matches the average value of \textit{Gaia} DR3, although adopting a conservative figure for its uncertainty. Given the large distance to the star, we corrected the photometry from interstellar extinction by means of the \textit{Gaia} DR3 estimates for the line-of-sight extinction in the $G$ band, $A_{G}$, and the reddening $E$($BP$-$RP$). Note this implementation is supported by the fact that our inferred value for the effective temperature is higher but comparable (within 200 \,K) to the \textit{Gaia DR3} estimate. The inputs for making our inference and the resulting fundamental parameters are presented in Table \ref{tab:star}. They convey that ASASSN-21qj is a main sequence star whose fundamental parameters resembles those of the Sun. 

\begin{table}[t!]
    \centering
    \caption{\textit{Gaia} DR3 photometric and astrometric values \citep{2021Gaia}, 
    stellar distance from \cite{2021BailerJones}, and our inferred fundamental parameters for ASASSN-21qj.}
    \begin{tabular}{lc}
        \hline\hline
        $\Pi$ ($mas$) & 1.763 $\pm$ 0.011\\
        $G$ ($mag$)     & 13.3708 $\pm$ 0.0028\\
        $A_{G}$ ($mag$) & 0.066$^{+0.005}_{-0.003}$\\ 
        $G_{BP}-G_{RP}$ ($mag$) & 0.815 $\pm$ 0.005\\
        $E$($G_{BP}-G_{RP}$) ($mag$) & 0.0354$^{+0.0024}_{-0.0019}$\\
        $M_{G}$ ($mag$) & 4.579$^{+0.014}_{-0.016}$\\       
        {[Fe/H]}    & -0.20 $\pm$ 0.20 \\
        \hline
        Distance (pc) & 556$^{+3}_{-4}$ \\
        Radius ($R_{\odot}$) & 0.977~$\pm$ 0.016  \\
        Effective temperature (K) & 5948~$\pm$~29 \\
        Mass ($M_{\odot}$) & 0.94~$\pm$~0.07 \\
        Luminosity ($L/L_{\odot}$) & 1.076$^{+0.007}_{-0.009}$ \\
        Bolometric magnitude 
        (mag) & 4.660$^{+0.018}_{-0.019}$ \\
        Surface gravity ($\log g$/$cgs$) & 4.43 $\pm$ 0.04 \\
        Age (Gyr) & 6~$\pm$~3 \\
        \hline
    \end{tabular}
    \label{tab:star}
\end{table}

\subsection{Lightcurve}

\begin{figure*}
    \includegraphics[width=\textwidth]{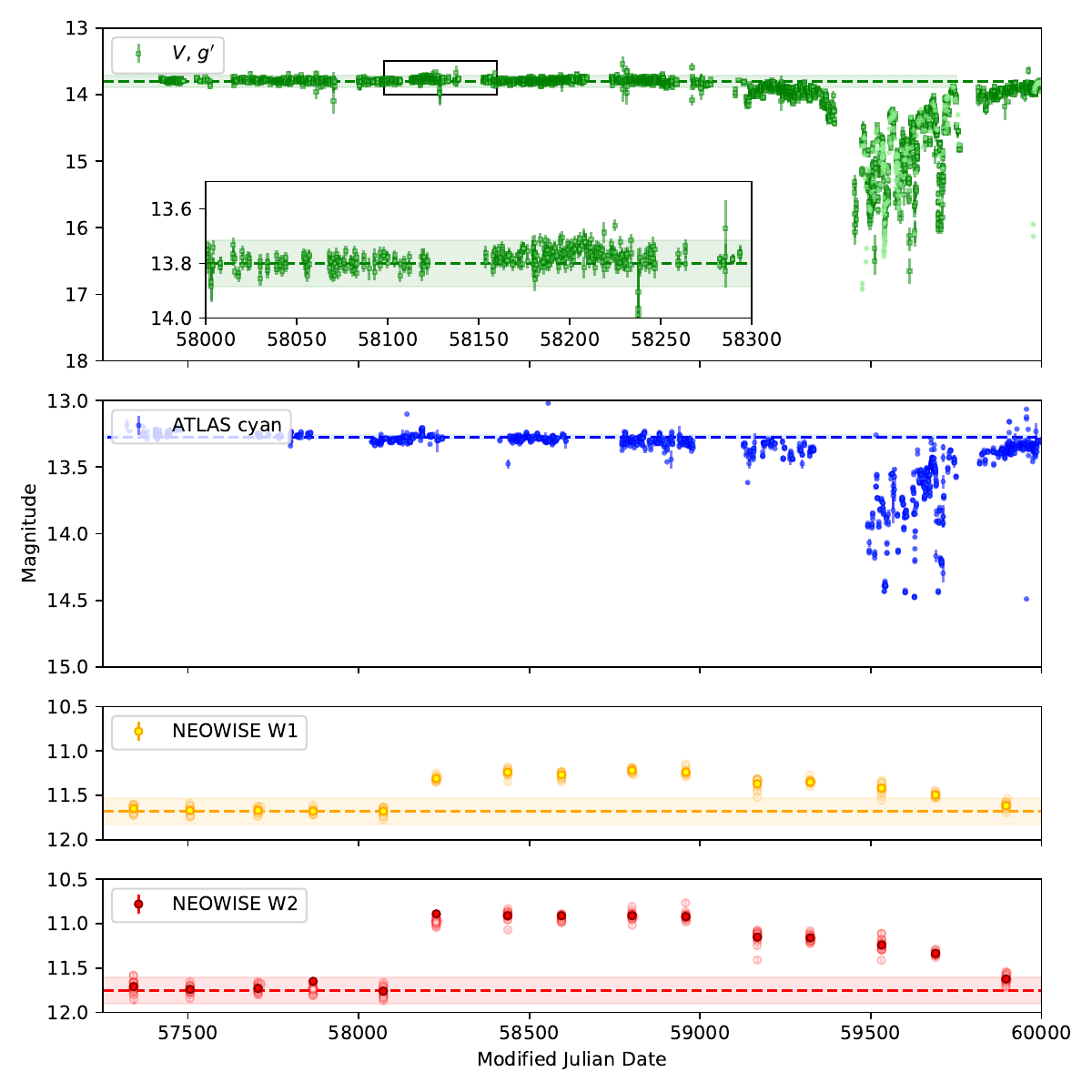}
    \caption{\textit{Top}: Optical photometry comprising ASAS-SN $V$ and $g^{\prime}$ measurements (dark green squares, NB: $V$ values were transposed to $g^{\prime}$ before plotting) and LCOGT $g^{\prime}$ (light green circles). The horizontal dashed line represents the unocculted $g^{\prime}$ magnitude of the star. An inset panel spans MJD 58000 to 58300, highlighting the region in between the two NEOWISE epochs shown in the lower two panels where the near-infrared excess appears. \textit{Upper Middle}: ATLAS `cyan' ($\lambda = 653$~nm) filter band photometry. The dashed line denotes the mean magnitude of the star in measurements prior to MJD 58500, whilst the shaded region denotes 3 times the mean uncertainty of those measurements. \textit{Lower Middle}: NEOWISE W1 (3.4~$\mu$m) measurements. Individual data points are translucent with the mean for each epoch presented as a solid data point. The shaded region denotes the $\pm~3-$sigma region around the AllWISE catalogue measurement, denoted by a dashed line. \textit{Bottom}: NEOWISE W2 (4.6~$\mu$m) measurements, presented as per the third panel.}
    \label{fig:timeseries}
\end{figure*}

\begin{figure}[!ht]
    \centering
    \includegraphics[width=0.48\textwidth]{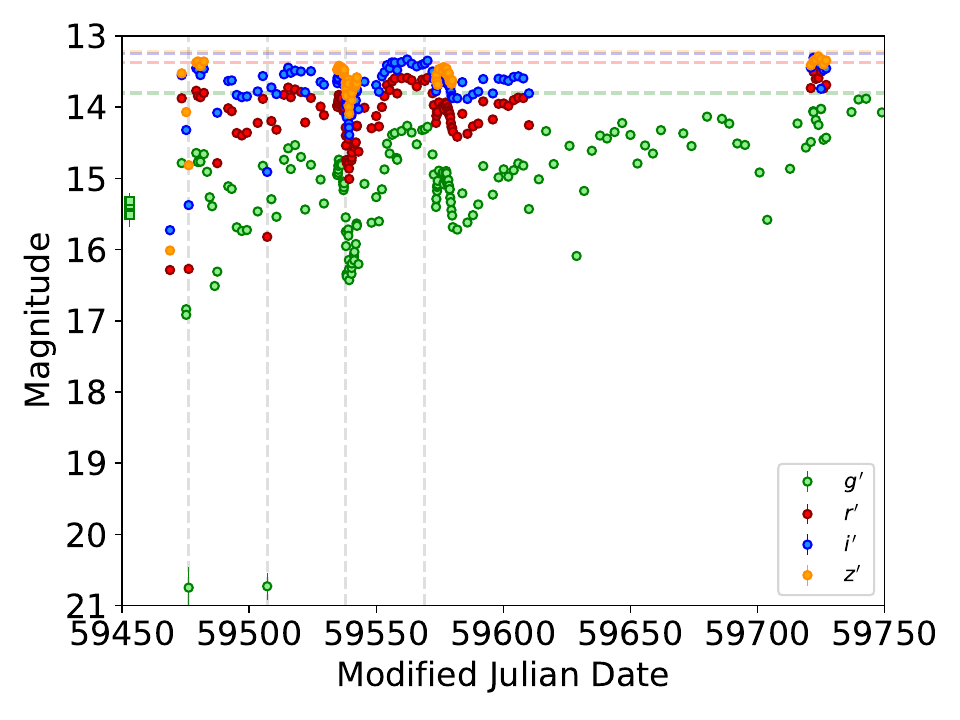}
    \caption{Close up of the ASASSN-21qj occultation events presented in Figure \ref{fig:timeseries} from MJD 59450 to 59750. Here we present multi-wavelength measurements comprising $g^{\prime}$ (green), $r^{\prime}$ (red), $i^{\prime}$ (blue), and $z$ (orange). The unocculted magnitude of the star in each filter is denoted by the dashed line in the respective colour. Vertical dashed grey lines denote the postulated period of 30.9 days based on the first two deep events, and their predicted recurrence.}
    \label{fig:timeseries_closeup}
\end{figure}

The lightcurve at optical wavelengths including ASAS-SN $V$ and $g^{\prime}$ measurements \citep{2014Shappee}, LCOGT $g^{\prime}$ (this work), and ATLAS `cyan' (653~nm) filter \citep{2018Tonry,2018Heinze}, and near-infrared wavelengths \citep[NEOWISE 3.4 and 4.6~$\mu$m;][]{neowise}  is presented in Figure \ref{fig:timeseries}, with a closeup of the multi-wavelength LCOGT observations given in Figure \ref{fig:timeseries_closeup}. The overall shape and duration of the dimming observed through the ASAS-SN/LCOGT and ATLAS observations are qualitatively similar, so we describe the event here with reference to the $g^{\prime}$ measurements from the former combined dataset. Around MJD 58200 a brief rise in the $g^{\prime}$ brightness is seen in ASAS-SN photometry, contemporaneously with the brightening at near-infrared wavelengths - this is marginal, but could be indicative of scattered light from the event that produced the dust responsible for the near-infrared excess and subsequent optical dimming. Prior to MJD 59000 no significant dimming of the star is observed in ASAS-SN monitoring, and the near-infrared excess remains constant. After that, a shallow decline in optical brightness is observed, as described in \cite{2021ATel14879....1R}, and a near-simultaneous decrease in the near-infrared excess is observed. The optical brightness then rapidly declines around 59400 shortly before the onset of LCOGT monitoring. From MJD 59450 we record a variety of structure in the occultation event at timescales of hours to days, broadly consisting of a series of deep, quasiperiodic dimming events superimposed on a monotonically declining occultation. 

The higher spatial resolution and sensitivity of LCOGT observations trace the dimming of below the 16 mag limit of ASAS-SN. At the onset of LCOGT observations we record a stellar magnitude of $g^{\prime} \simeq~$17~mag. Early in the evolution of the occultations two very deep dimming events were observed on MJDs 59476 and 59507 reducing the star to $g^{\prime} \simeq 20.3~$mag. The likelihood of two such deep events occurring within a single occultation being unconnected seems unlikely given the ongoing, decreasing overall dimming of the star in the intervening period. If we therefore assume that this is a recurrence of the same dense clump transiting the star, we infer an orbital period of 30.9~$\pm$~0.5 days for the material responsible, equivalent to a semi-major axis of 0.19~$\pm$~0.04~au.

Based on this potential period, additional observations were scheduled around the next couple of times this clump was expected to occult the star at MJDs 59537 and 59569. For the first of these, the expected occultation was a day late, demonstrating the clump was not periodic, and at most quasiperiodic. Whilst later than predicted, a relatively deep occultation did occur reducing the star to $g^{\prime} > 16~$mag, from a baseline of 14~mag either side of this dip. From this we concluded that the structure of the clump was evolving, becoming larger and more diffuse over time. At the second predicted epoch we found a delay of 4 days from the predicted time. The clump had significantly broadened only dimming the star to $g^{\prime} > 15.5~$mag with two distinct dips within the overall event. The overall structure had changed to span several days with an extended tail characteristic of a comet-like transit before the star resumed its pre-dip brightness. From this we concluded that the original spatially compact clump could be fragmenting into several distinct sub-clumps, or involve multiple bodies that could now be discerned due to their increasing separation. Following this, another two deep occultations are recorded in the sparser $g^{\prime}$ monitoring following MJD 59600; these follow the established pattern of being increasingly extended in time and shallower in dimming, although the measurements are too sparse (a cadence of every other day) to identify any substructure.

\subsection{Periodicity}

\begin{figure*}
    \centering
    \includegraphics[width=\textwidth]{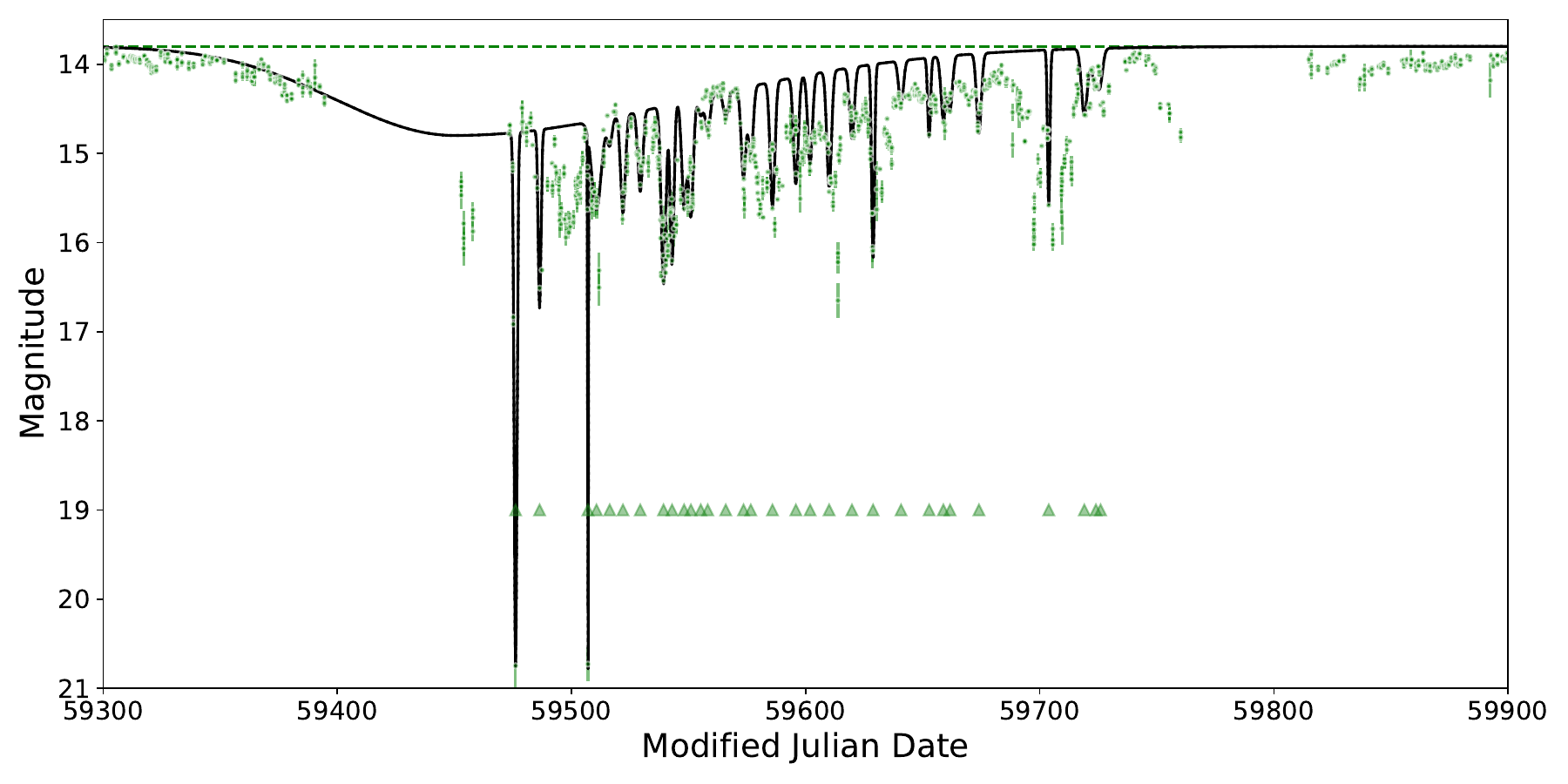}
    \caption{ASAS-SN and LCOGT $g^{\prime}$ photometry of ASASSN-21qj between MJD 59300 and 59900. The observations are denoted by green data points with 1-$\sigma$ uncertainties. Over-plotted on the observations is a two component model of the obscuration. A broad, asymmetric and shallow component is centered at MJD 59450. Narrow, deeper events are superimposed upon that (35 in total), with the location of narrow dimming events in the time series denoted by the upward pointing green triangles. These events were identified by searching for observations that were 3-$\sigma$ fainter than the preceeding and subsequent data point in the time series, taking into account the underlying broad, shallow obscuration component. This model does not capture, and does not seek to capture, the full complex behaviour of the dimming event, where it is clear that intermediate duration dimming events lasting several days to tens of days occurred. Rather, it is illustrative of the extremes of the timescales involved within the overall dimming event for ASASSN-21qj.} \label{fig:dimming}
\end{figure*}

The occultation of ASASSN-21qj can be described as the conflation of two effects, namely a broad, low level extinction covering the star between MJDs 59000 and 60000 that is interspersed with multiple short duration events lasting a few days to a few tens of days. The light curve is shown in Figure \ref{fig:dimming} with the broad and narrow dimming events denoted by triangles. We seek to disentangle these two effects on the stellar brightness in order to determine the distribution of depths for narrow events that are superimposed on the broader overall dimming.

We model the broad extinction as an asymmetric Gaussian scaled to the shallow ingress and egress at the wings of the extinction event. The shape of this broad profile is described by a date of peak dimming $t_{\rm peak}$, peak depth $A_{\rm peak}$, and standard deviations before and after the peak $\sigma_{\rm prior}$ and $\sigma_{\rm post}$. The broad, shallow component of the obscuration has a maximum depth of 1 magnitude at MJD 59450 and standard deviations of 50 days before the peak, and 100 days after the peak. This asymmetric shape is characteristic of transiting dust clouds associated with exocomets.

Narrow, deep events in the time series are identified by searching for data points for which the preceding and subsequent data points are brighter by 3-$\sigma$, these events are modelled as symmetric Gaussians in similar fashion to the broad component. The peak depths of the narrow events are determined with reference to the level of broader extinction occurring at the same time. We identify 35 narrow dimming events in the $g^{\prime}$ light curve. During high cadence observations we see strong variability in the extinction within a single night's photometry. We are thus insensitive to rapid changes in the structure of the cloud mainly comprising smaller dimming events. Likewise, the adopted method to identify narrow dimming events does not track broader, shallow events as can be seen by the lack of dimming events identified in the lightcurve during egress from overall extinction beyond MJD 59800. The estimated number of occultation events is therefore a lower limit to the total number.

We also examine the complete light curve using the Lomb Scargle periodogram as implemented in {\sc astropy} to search for significant periods. Combining the $g^{\prime}$ measurements from LCOGT and ASAS-SN we obtain a peak period at 19.6~days, with a false alarm probability $< 1\%$. Examining the data sets separately, we find peak periods of 29.9 days for the ASAS-SN observations, and 41.6 days for the LCOGT observations, both with false alarm probabilities $< 1\%$. The LCOGT periodogram is noisy with several substantial aliases due to the irregular sampling and shorter time span of the observations compared to the ASAS-SN periodogram. Taking the peak period from the combined periodogram is therefore suspect. At the very least we can state there is good evidence for periodic behaviour between 19 and 42 days present in the light curve of ASASSN-21qj, consistent with the detected infrared excess.

The presence of a 31 day quasi-periodicity was previously inferred based on two deep dimming events early in the occultation (see Figures \ref{fig:timeseries_closeup} and \ref{fig:dimming}). Combined with the speculative periodic behaviour seen here and the presence of near-infrared excess indicative of close-in dust, we can estimate the size and structure of dust clumps responsible for the narrow, deep events within the occultation. Assuming the dust lies on circular orbits with periods between 19 and 42 days (equivalent to a semi-major axes from 0.14 to 0.23 au), the transiting velocity of the dust would be 74 to 60 km/s. The two deepest, narrowest features have total durations of 5.4 days, equivalent to a cloud extent along the orbit of 0.23 to 0.19 au. Given the near complete extinction of the stellar flux during these events, the cloud must also have a substantial vertical extent, with a scale height around 0.03 to completely cover the stellar disc.

\subsection{Colour-colour plots}

Colour-colour plots ($g^{\prime}-r^{\prime}$ vs $r^{\prime}-i^{\prime}$ and $g^{\prime}-r^{\prime}$ vs $r^{\prime}-z^{\prime}$) are used to determine the reddening vectors associated with the source extinction from the LCOGT time series photometry. We use the computer code {\sc optool} \citep{2021Dominik} to calculate the $Q_{\rm ext} (= Q_{\rm abs} + Q_{\rm sca})$ values for a range of dust grain compositions and two size distributions. We used the in-built optical constants from {\sc optool} to generate the $Q_{\rm ext}$ values for materials including astronomical silicate \citep{2003Draine}, various crystalline and amorphous iron-magnesium silicates \citep{1995Dorschner,1998Jaeger,2001Fabian,2006Suto}, carbon-bearing species \citep{1993Draine,1996Henning}, quartz \citep{2007Kitamura}, and iron particles \citep{1996Henning}. 

We consider two grain size distributions: 1) a power law ($dN = a^{-q} da$) between a minimum (free) and maximum grain size (fixed as 1~mm, typical for debris discs) with a fixed size distribution exponent of 3.5 appropriate for a collisional cascade \citep{1969Dohnanyi} and 2) a log-normal with a width of 0.5 and a free peak size, considering contributions from sizes between 0.01 and 10~$\mu$m. In each case, the free parameter was allowed to vary between 0.1 and 1.1~$\mu$m in steps of 0.01~$\mu$m when calculating $Q_{\rm ext}$, determined from the complex optical constants assuming Mie theory. For a Sun-like star the typical blowout size for a dust grain is $\simeq$0.5~$\mu$m \citep{2010Krivov}, so we include transient dust with this range of grain sizes. The optical depth $\tau$ and extinction of a dust cloud which fully covers the stellar disc are then calculated from the $Q_{\rm ext}$ values for each combination of composition and size distribution, which are then compared to the measured reddening vectors by least-squares fit to identify those which best match the observations. 

The colour-colour plots for some of the materials tested in this analysis are presented in Figures \ref{fig:col_col_ln} and \ref{fig:col_col_pl}, along with the best-fit material and size from the modelling in Figure \ref{fig:col_col_bf}. We find adequate fits to the reddening curves for {several materials with both power law and log-normal size distributions. For both size distributions, a wide variety of grain sizes and materials can be matched to $g^{\prime}-r^{\prime}$ vs $r^{\prime}-i^{\prime}$ colour, but only a handful provide a good match to both $g^{\prime}-r^{\prime}$ vs $r^{\prime}-i^{\prime}$ and $g^{\prime}-r^{\prime}$ vs $r^{\prime}-z^{\prime}$ simultaneously. For both size distributions the best fit composition was amorphous pyroxene (Mg$_{1.6}$Fe$_{0.4}$Si$_{2}$O$_{6}$). For the power law size distribution the best fit minimum grain size was 0.29$^{+0.01}_{-0.18}~\mu$m, whereas for the log-normal size distribution the best fit peak size was 0.88$^{+0.05}_{-0.03}~\mu$m. If we favour the power law distribution, the presence of small, sub-blowout dust grains indicates ongoing collisional grinding of larger bodies. The dissipation of the occulting screen of material during the event between dips would also be consistent with the rapid removal of these grains.

\begin{figure*}
    \centering
    \includegraphics[width=0.32\textwidth]{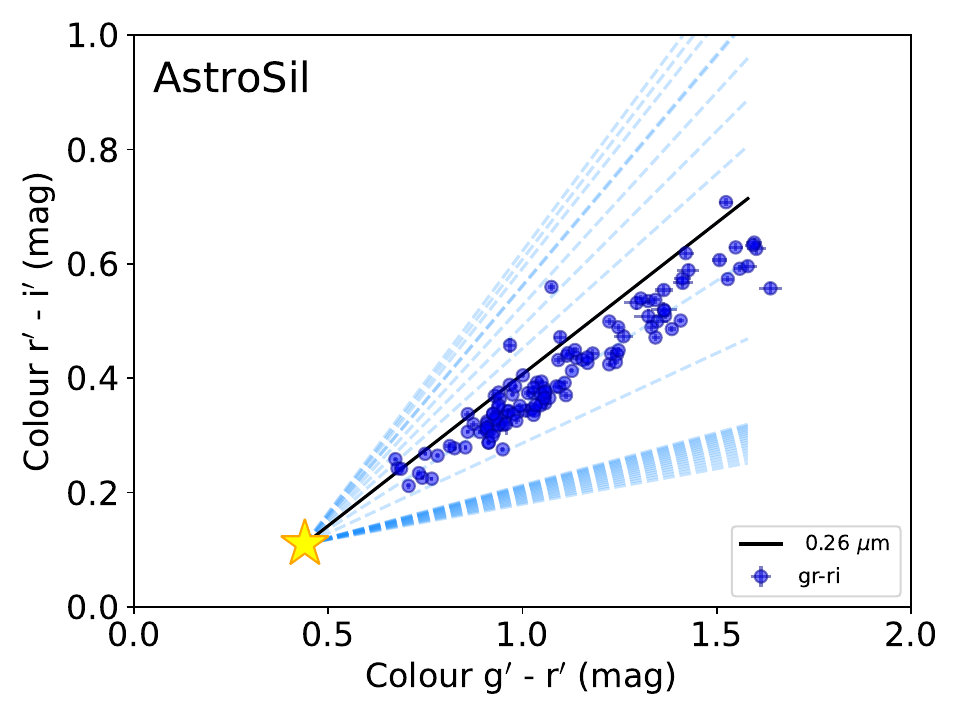}
    \includegraphics[width=0.32\textwidth]{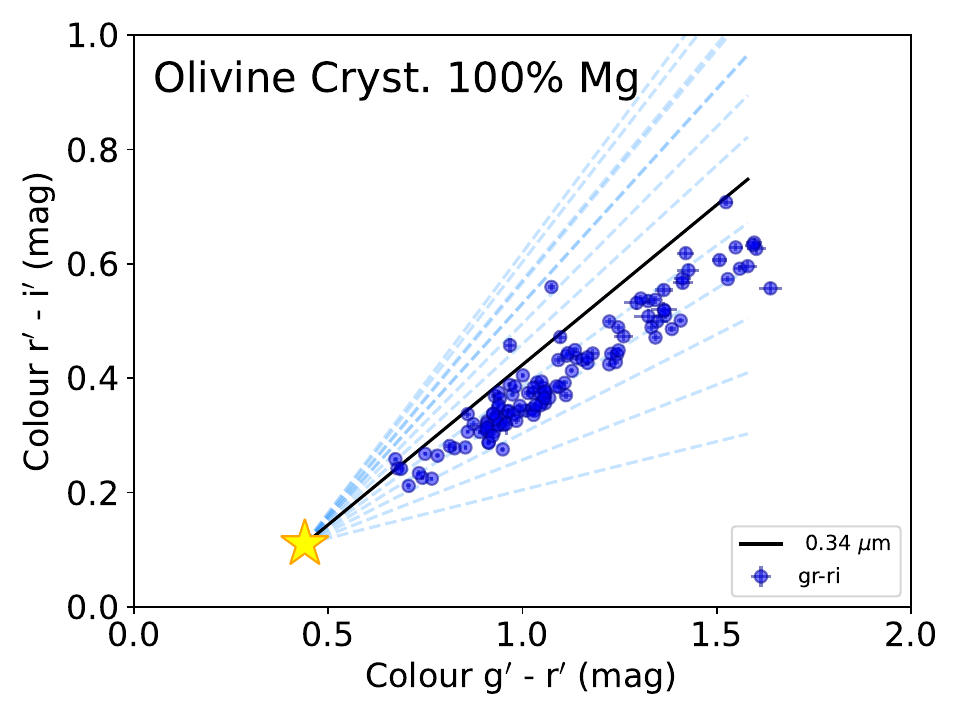}
    \includegraphics[width=0.32\textwidth]{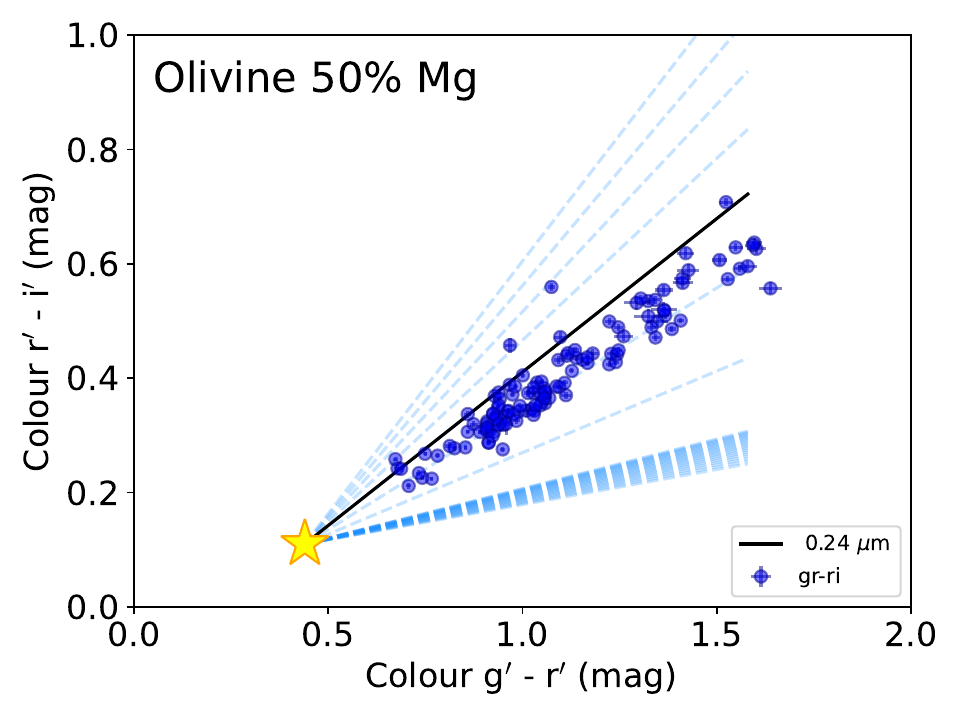}
    \includegraphics[width=0.32\textwidth]{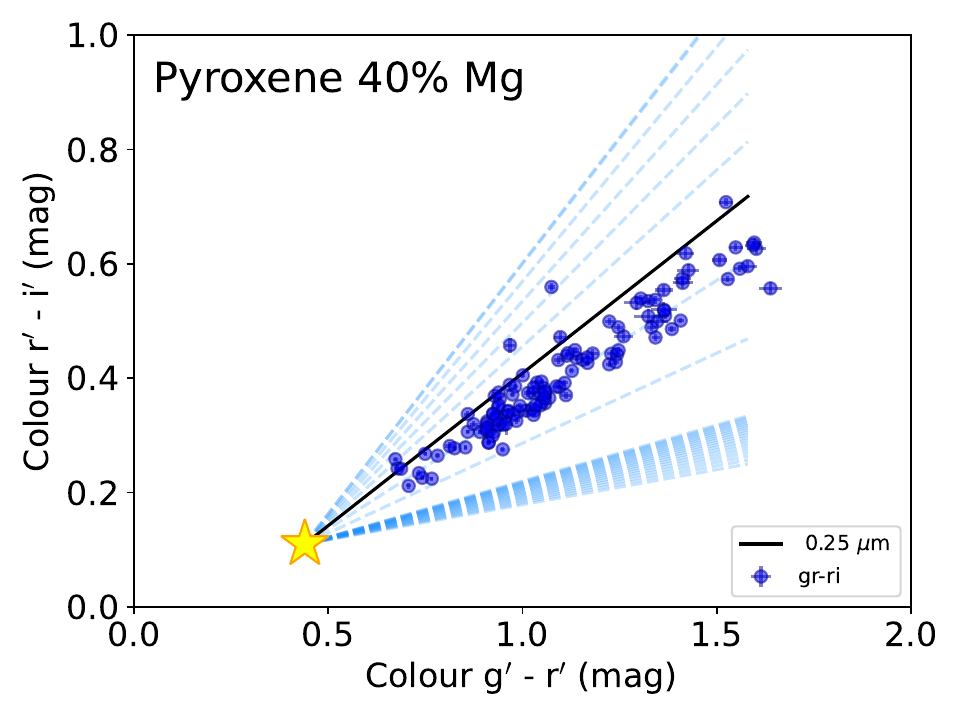}
    \includegraphics[width=0.32\textwidth]{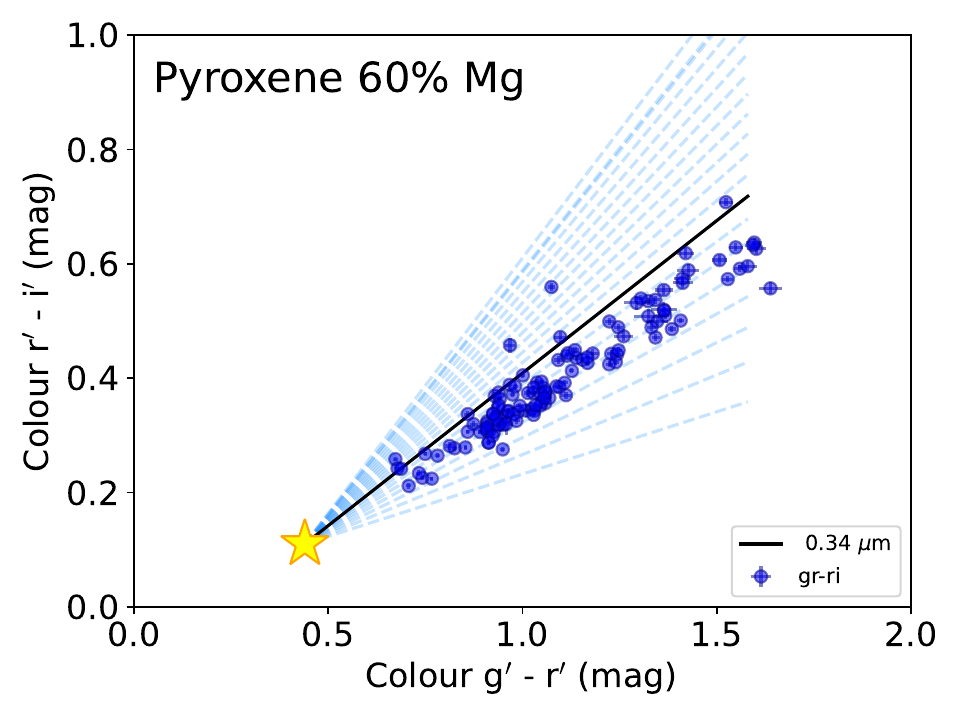}
    \includegraphics[width=0.32\textwidth]{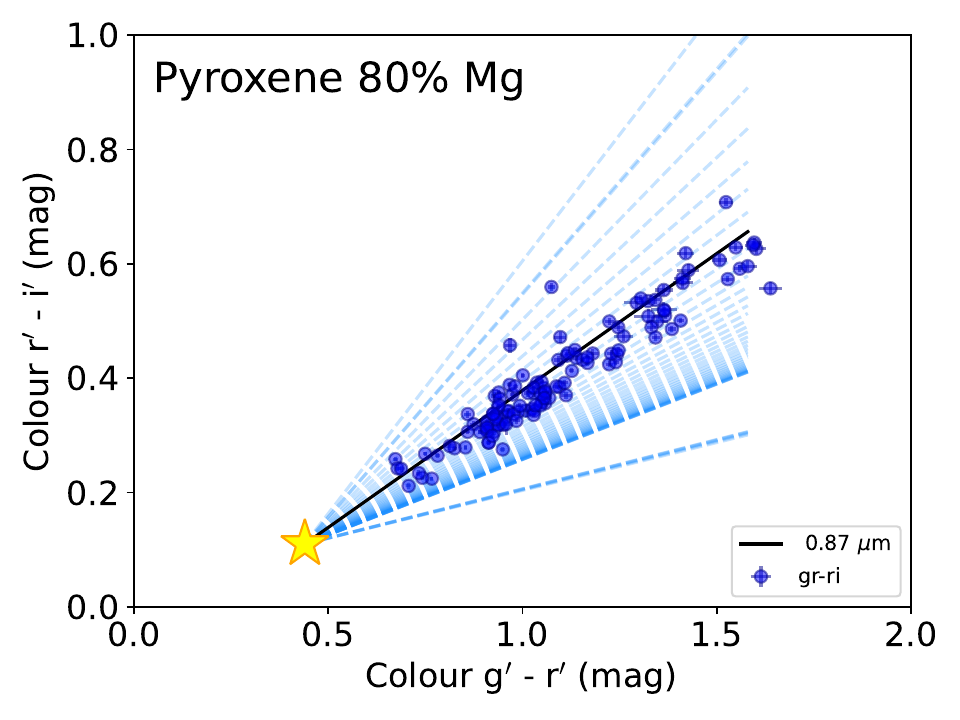}
    \includegraphics[width=0.32\textwidth]{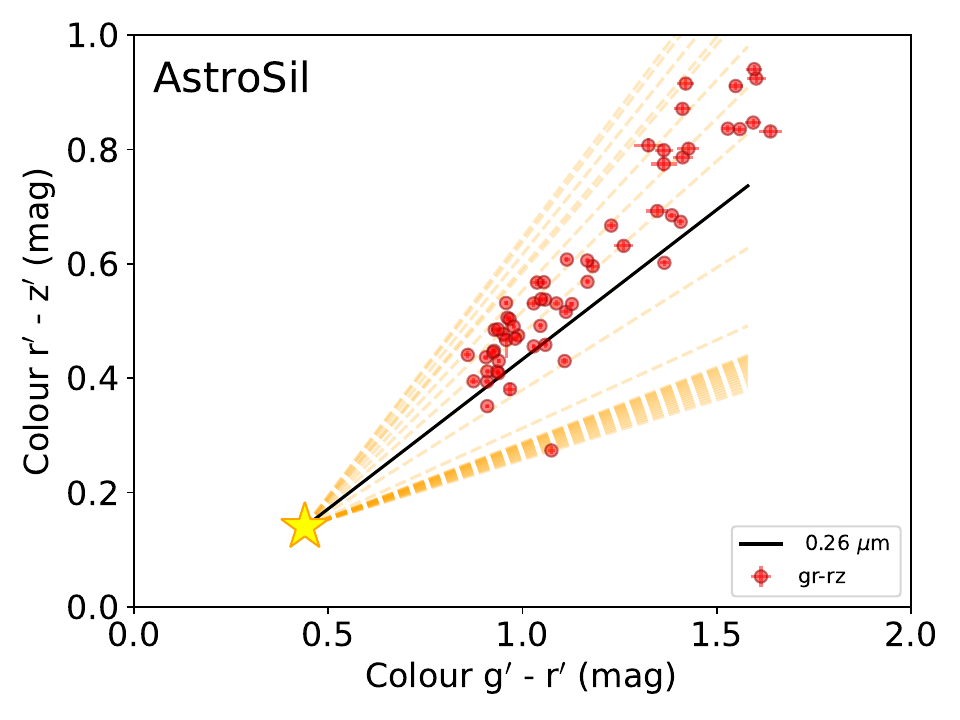}
    \includegraphics[width=0.32\textwidth]{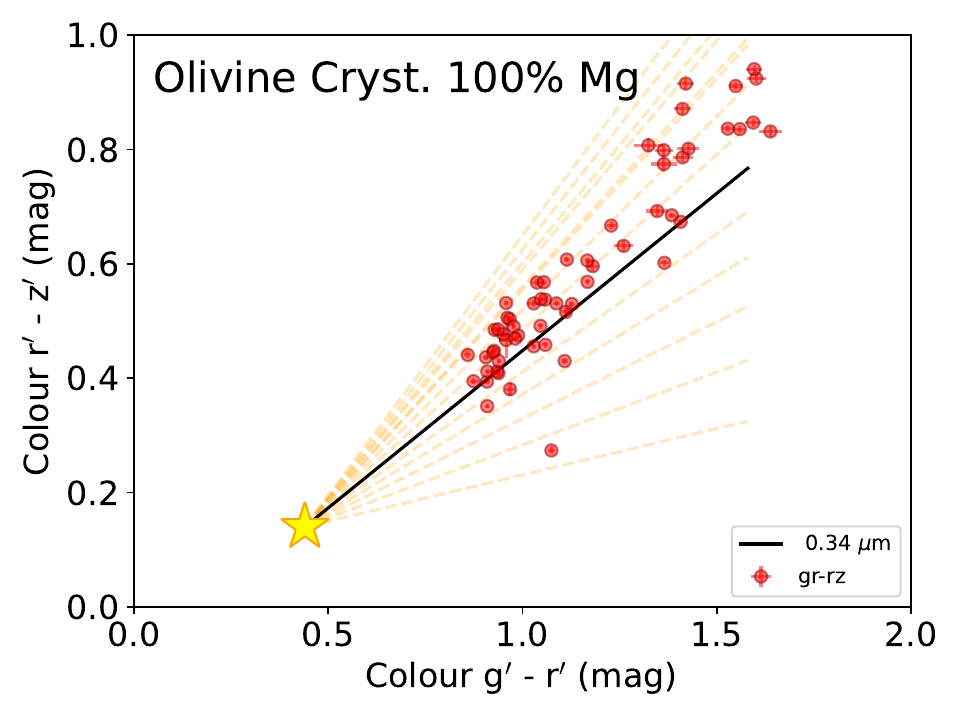}
    \includegraphics[width=0.32\textwidth]{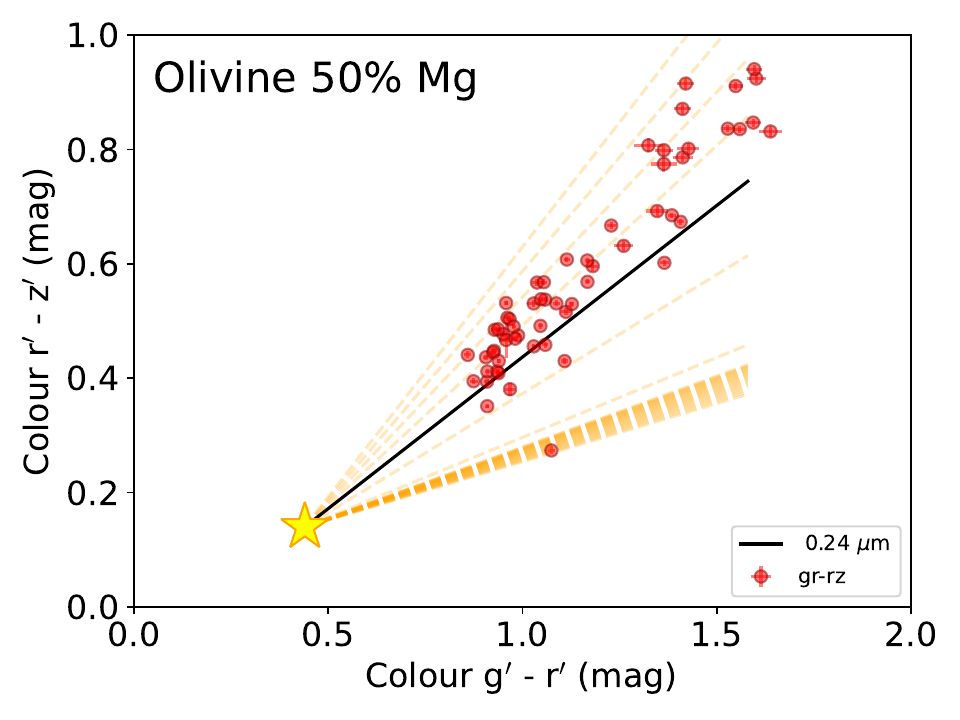}
    \includegraphics[width=0.32\textwidth]{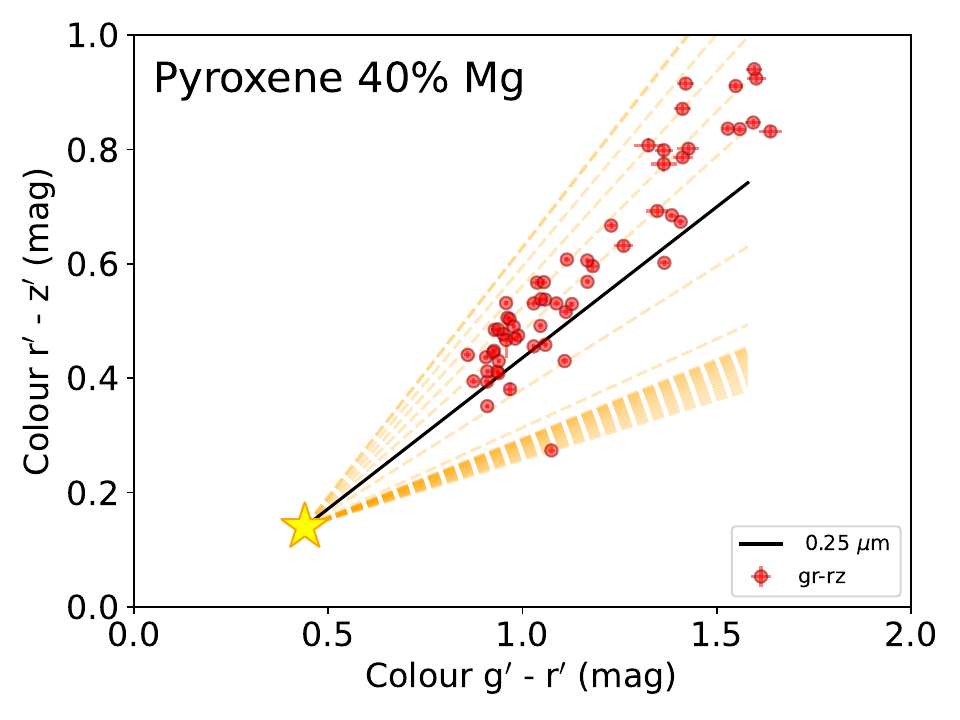}
    \includegraphics[width=0.32\textwidth]{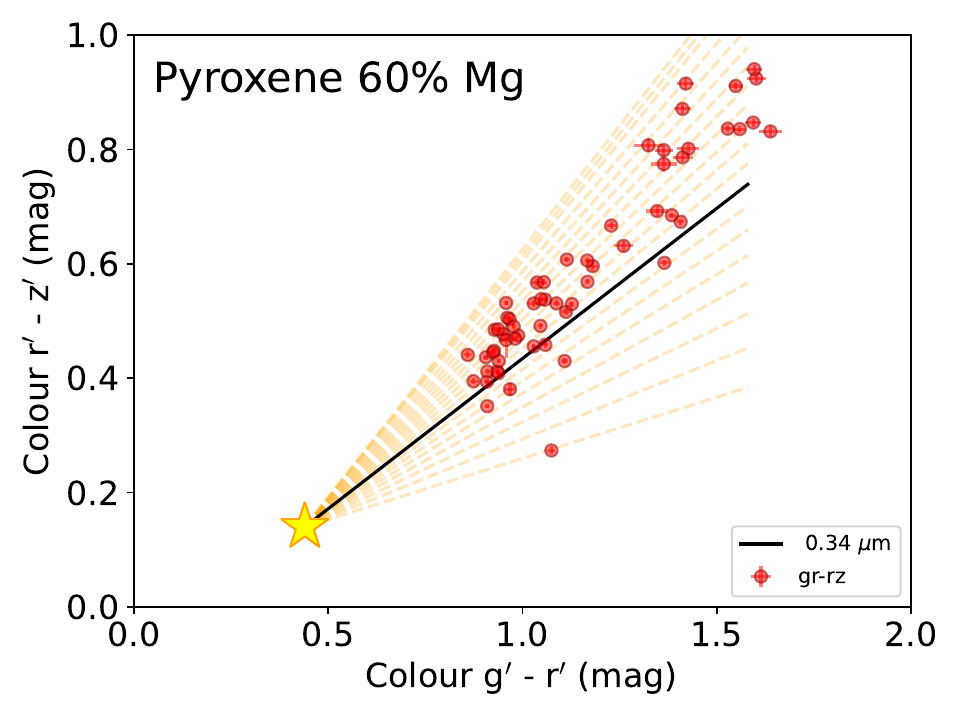}
    \includegraphics[width=0.32\textwidth]{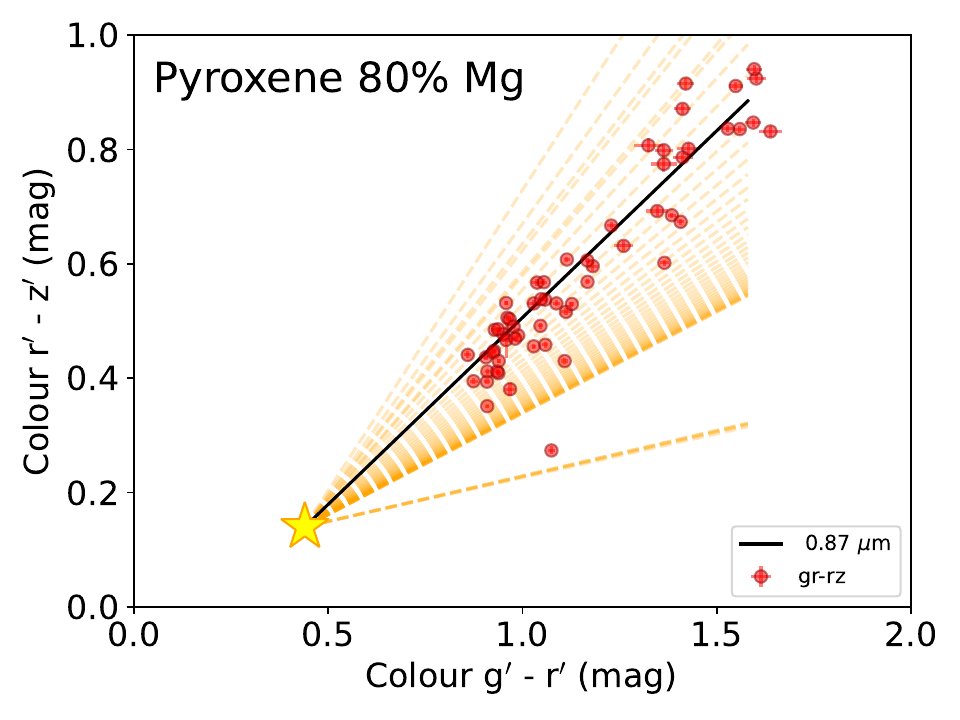}
    \caption{Colour-colour plots of ASASSN-21qj using the log-normal size distribution to calculate reddening vectors for a dust cloud occulting the star. Six materials out of the twelve materials tested are presented in this figure. The top two rows show the $g^{\prime}-r^{\prime}$ vs $r^{\prime}-i^{\prime}$ plots, whilst the bottom two rows show the $g^{\prime}-r^{\prime}$ vs $r^{\prime}-z^{\prime}$ plots. The LCOGT colours are presented as blue or orange data points, with the unreddened colour expected of a G2~V star shown as the orange '$\star$' symbol. Blue or orange dashed lines denote the reddening vectors for distributions of grains with different peak sizes between 0.1 and 1.2~$\mu$m consistent with the observations. The solid line denotes the vector for the best fitting size for each material taking into account both the $g^{\prime}-r^{\prime}$ vs $r^{\prime}-i^{\prime}$ and $g^{\prime}-r^{\prime}$ vs $r^{\prime}-z^{\prime}$ colours.}
    \label{fig:col_col_ln}
\end{figure*}

\begin{figure*}
    \centering
    \includegraphics[width=0.32\textwidth]{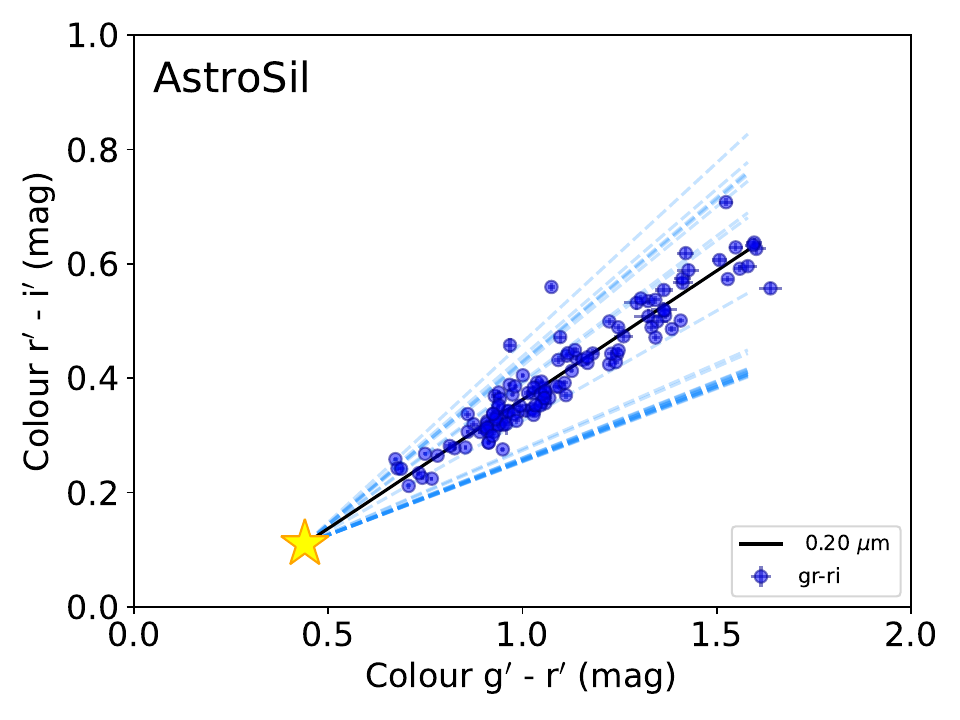}
    \includegraphics[width=0.32\textwidth]{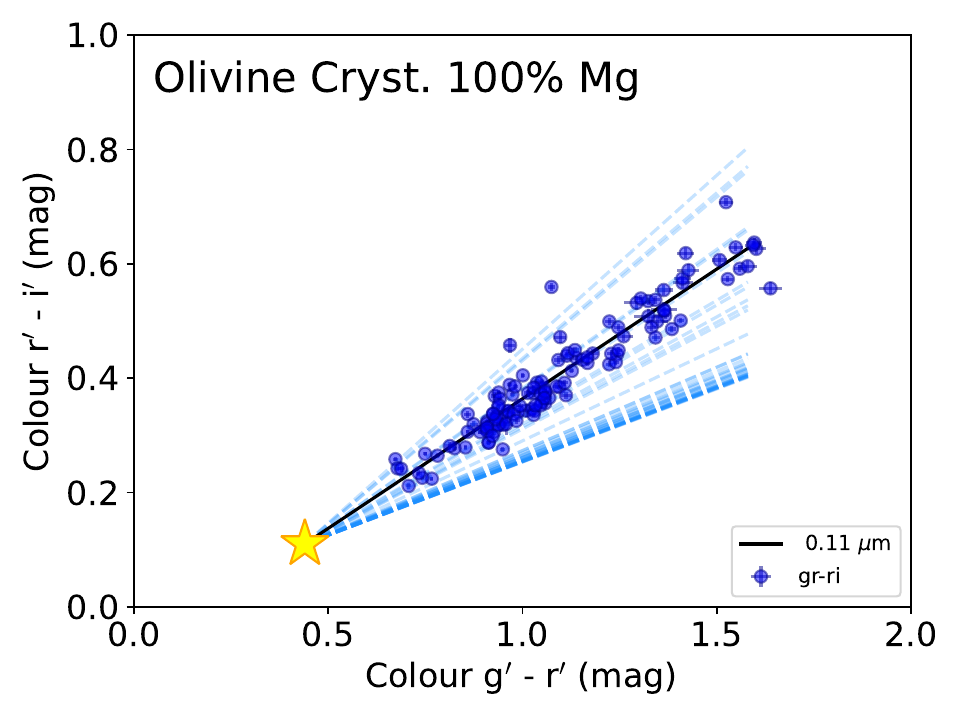}
    \includegraphics[width=0.32\textwidth]{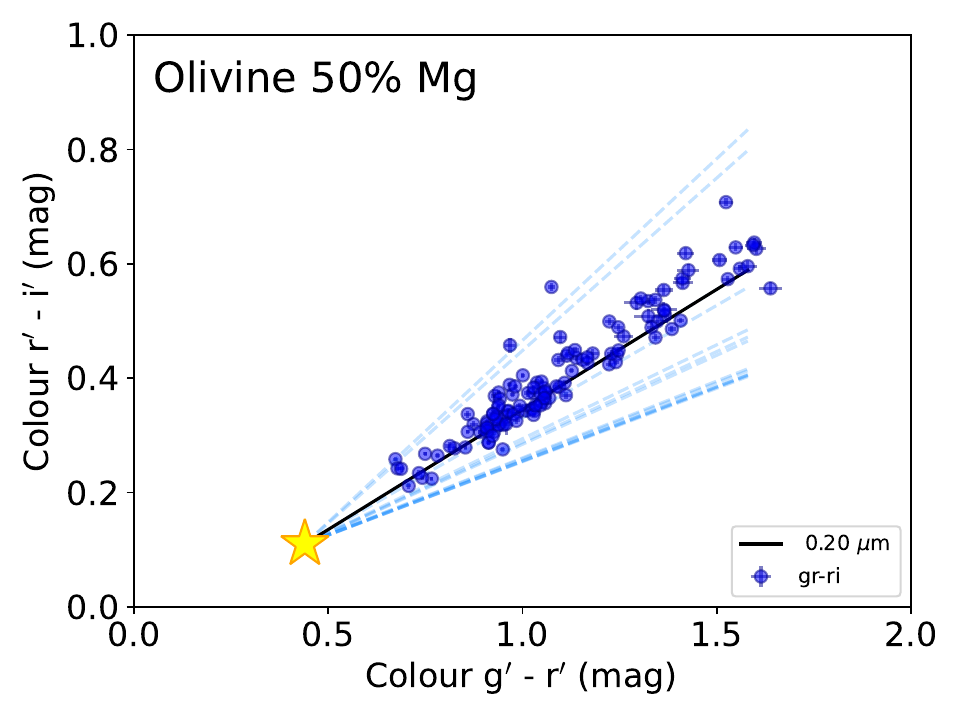}
    \includegraphics[width=0.32\textwidth]{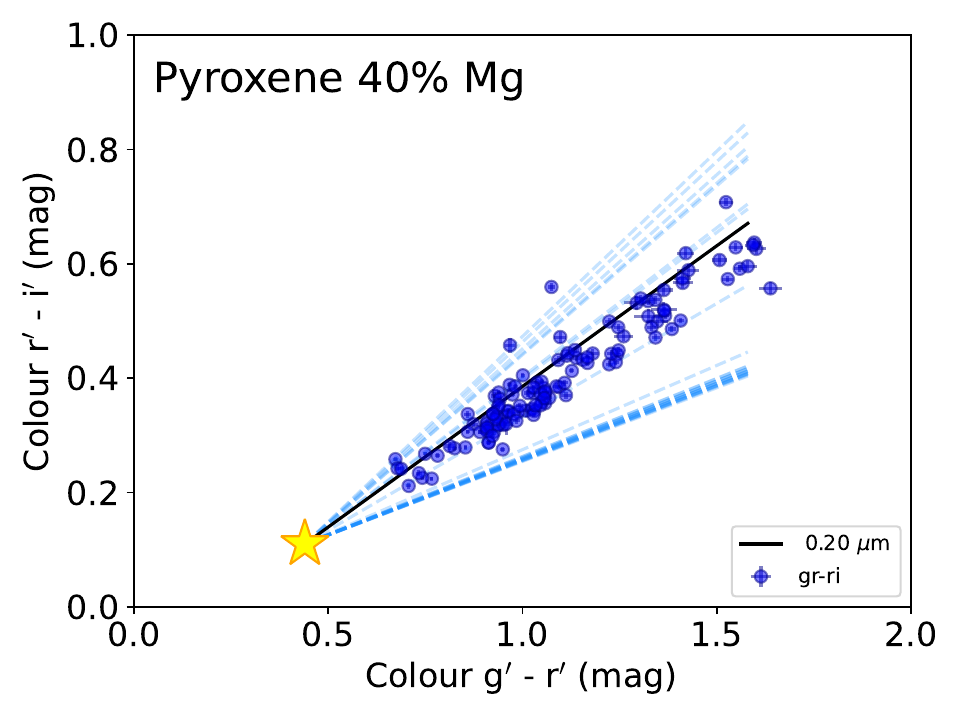}
    \includegraphics[width=0.32\textwidth]{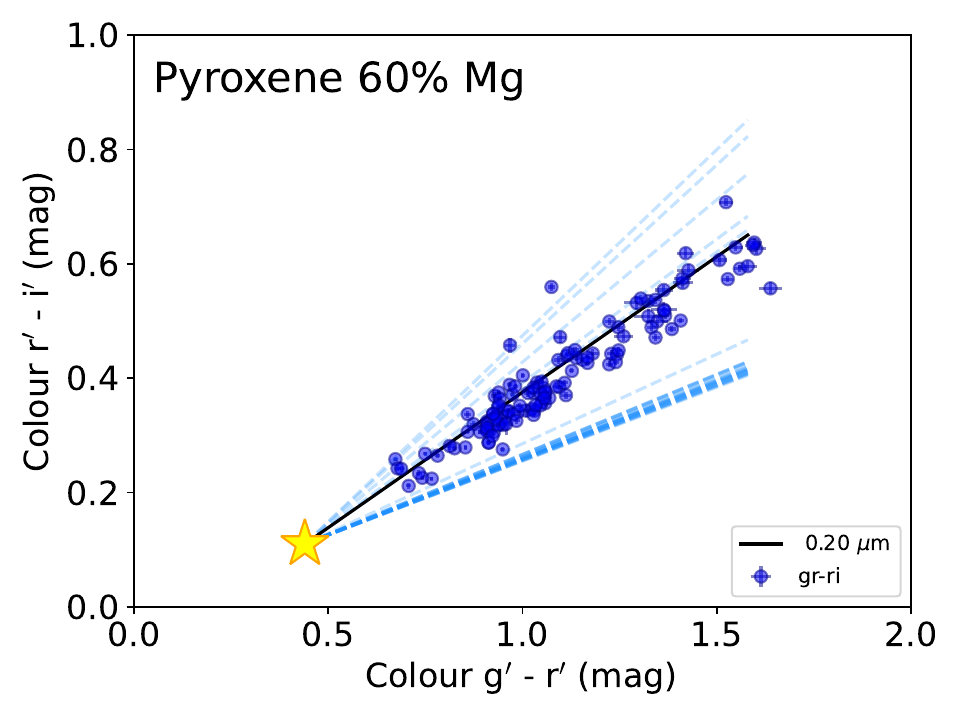}
    \includegraphics[width=0.32\textwidth]{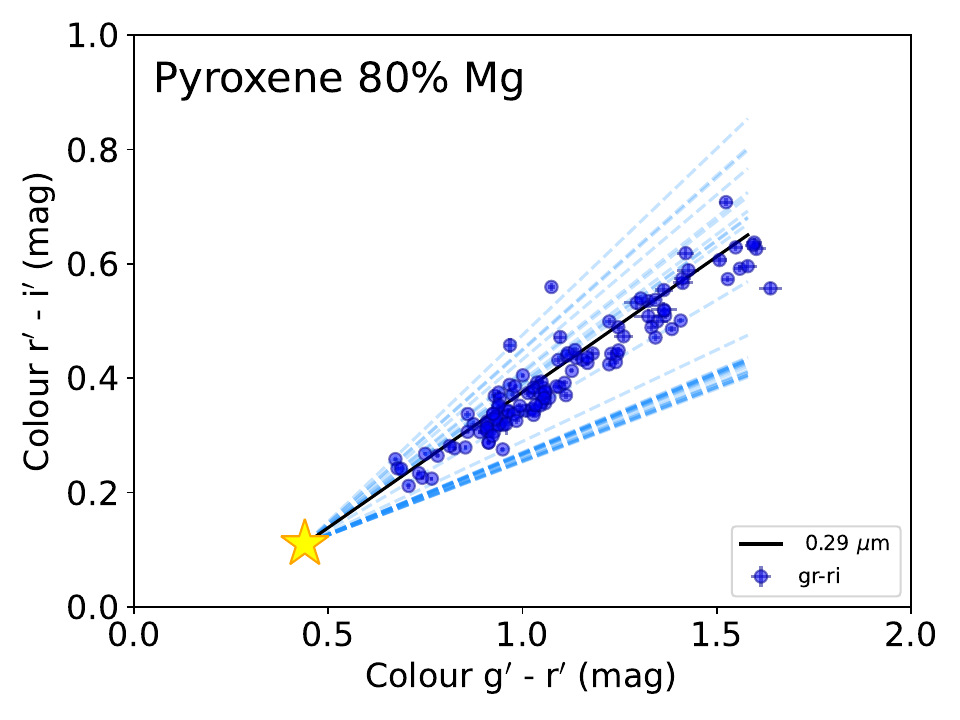}
    \includegraphics[width=0.32\textwidth]{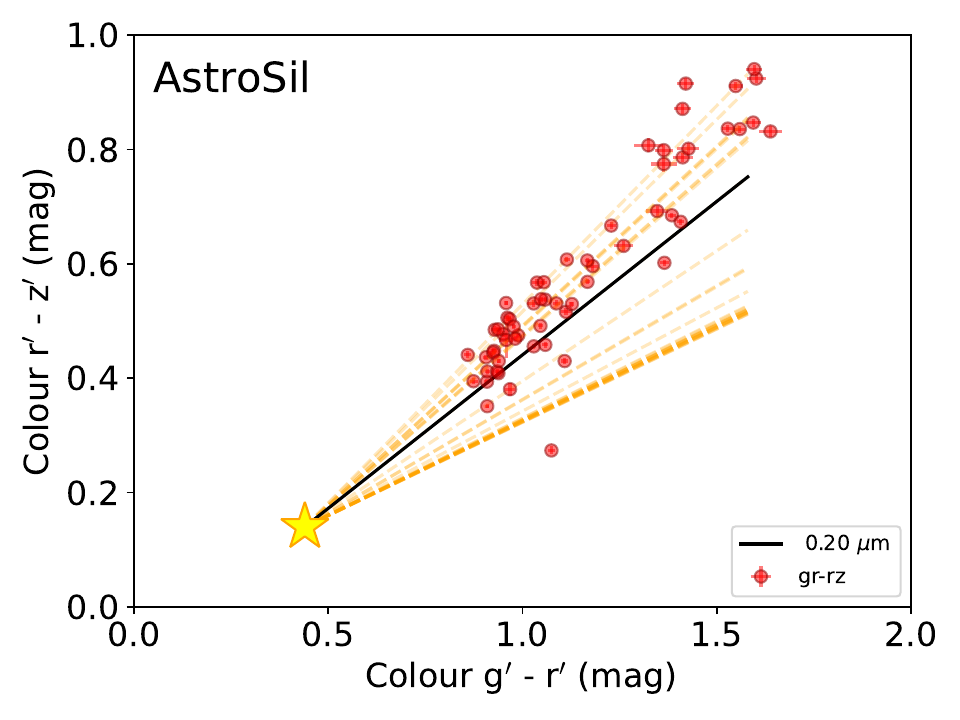}
    \includegraphics[width=0.32\textwidth]{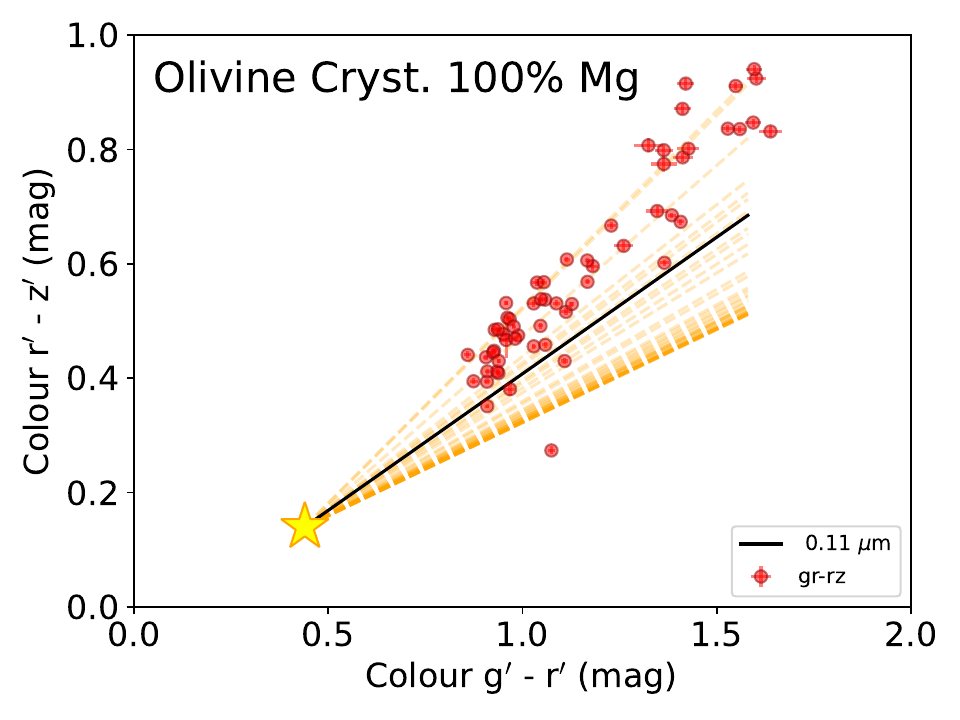}
    \includegraphics[width=0.32\textwidth]{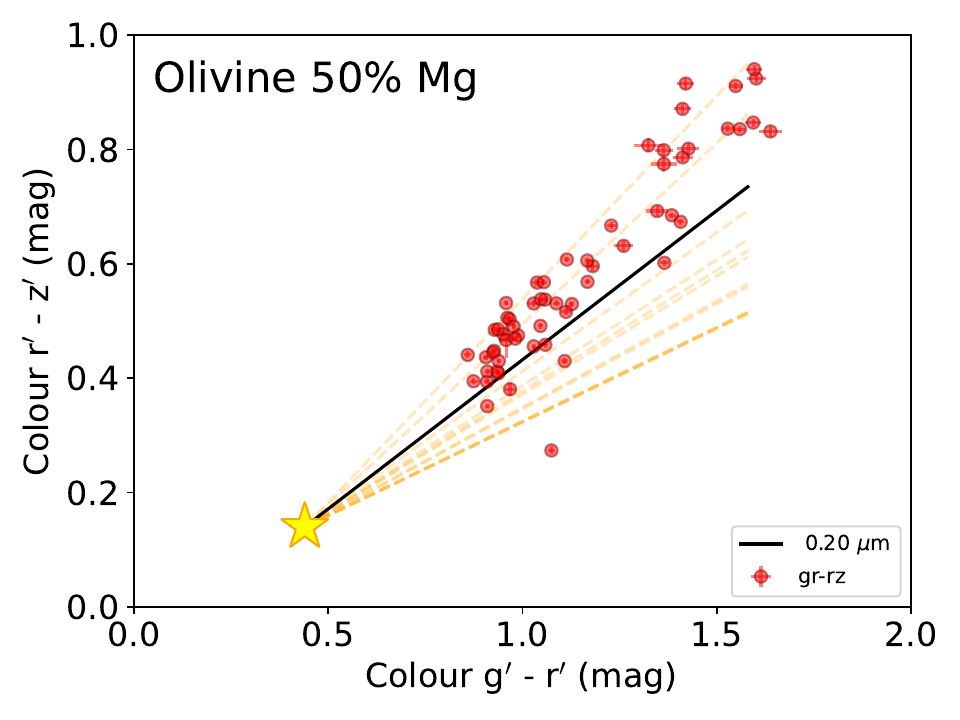}
    \includegraphics[width=0.32\textwidth]{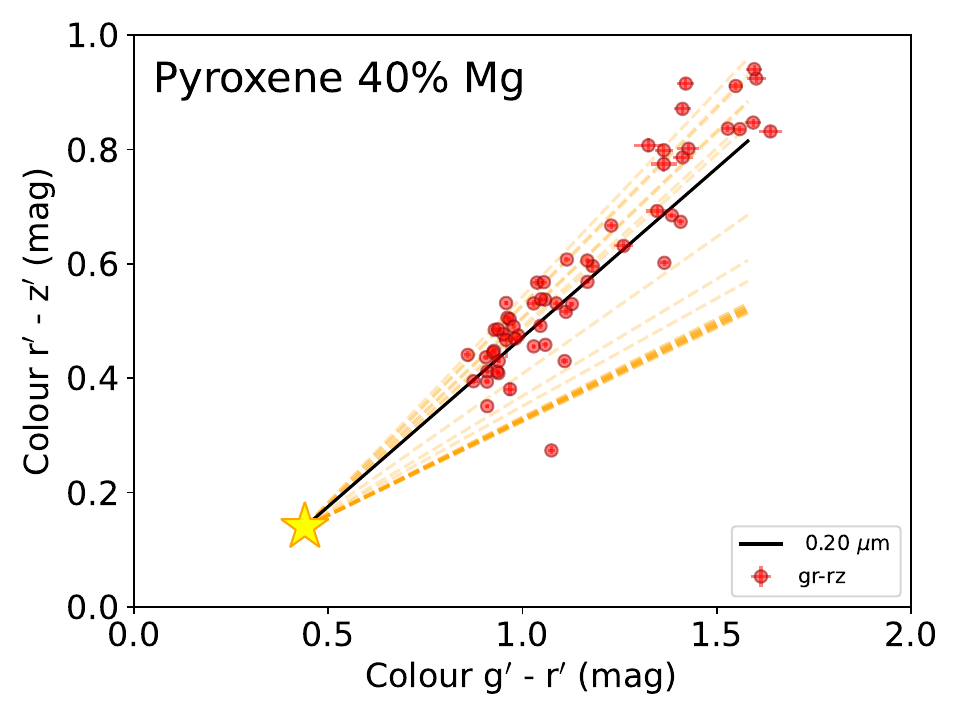}
    \includegraphics[width=0.32\textwidth]{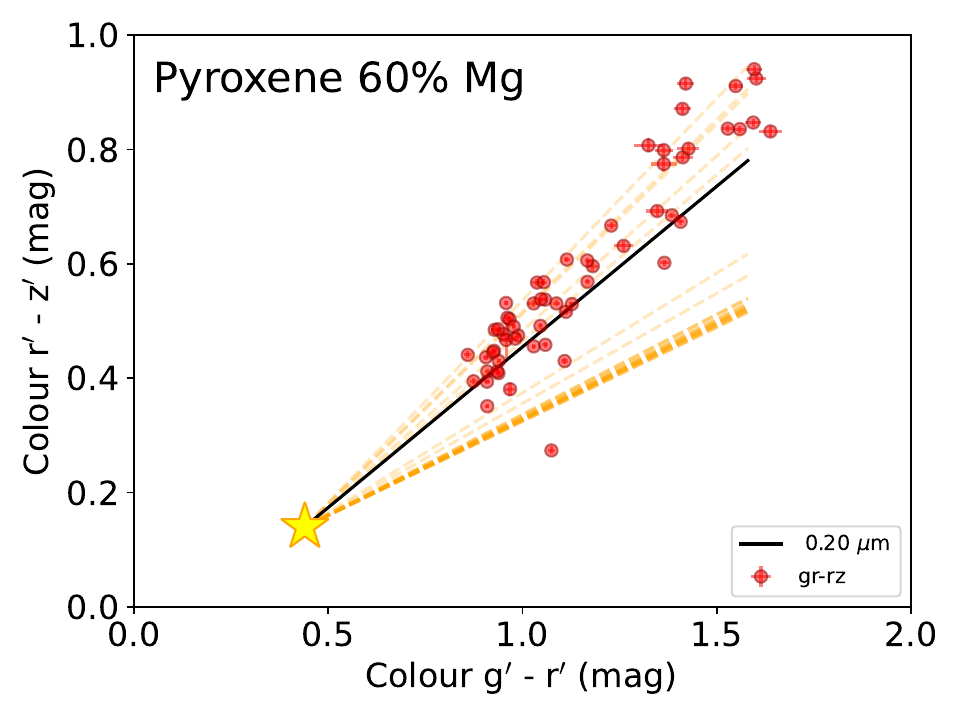}
    \includegraphics[width=0.32\textwidth]{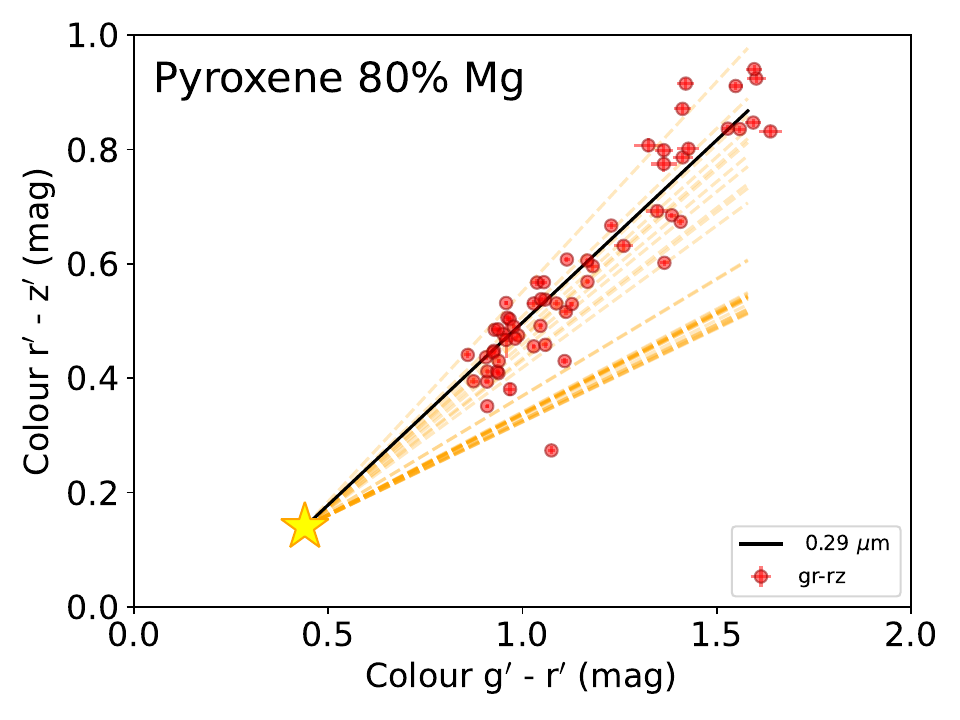}
    \caption{Colour-colour plots of ASASSN-21qj using the power law size distribution to calculate reddening vectors for a dust cloud occulting the star. Six materials out of the twelve materials tested are presented in this figure. The top two rows show the $g^{\prime}-r^{\prime}$ vs $r^{\prime}-i^{\prime}$ plots, whilst the bottom two rows show the $g^{\prime}-r^{\prime}$ vs $r^{\prime}-z^{\prime}$ plots. The LCOGT colours are presented as blue or orange data points, with the unreddened colour expected of a G2~V star shown as the orange '$\star$' symbol. Blue or orange dashed lines denote the reddening vectors for distributions of grains with different minimum sizes between 0.1 and 1.2~$\mu$m consistent with the observations. The solid line denotes the vector for the best fitting size for each material taking into account both the $g^{\prime}-r^{\prime}$ vs $r^{\prime}-i^{\prime}$ and $g^{\prime}-r^{\prime}$ vs $r^{\prime}-z^{\prime}$ colours.}
    \label{fig:col_col_pl}
\end{figure*}

\begin{figure*}
    \centering
    \includegraphics[width=0.48\textwidth]{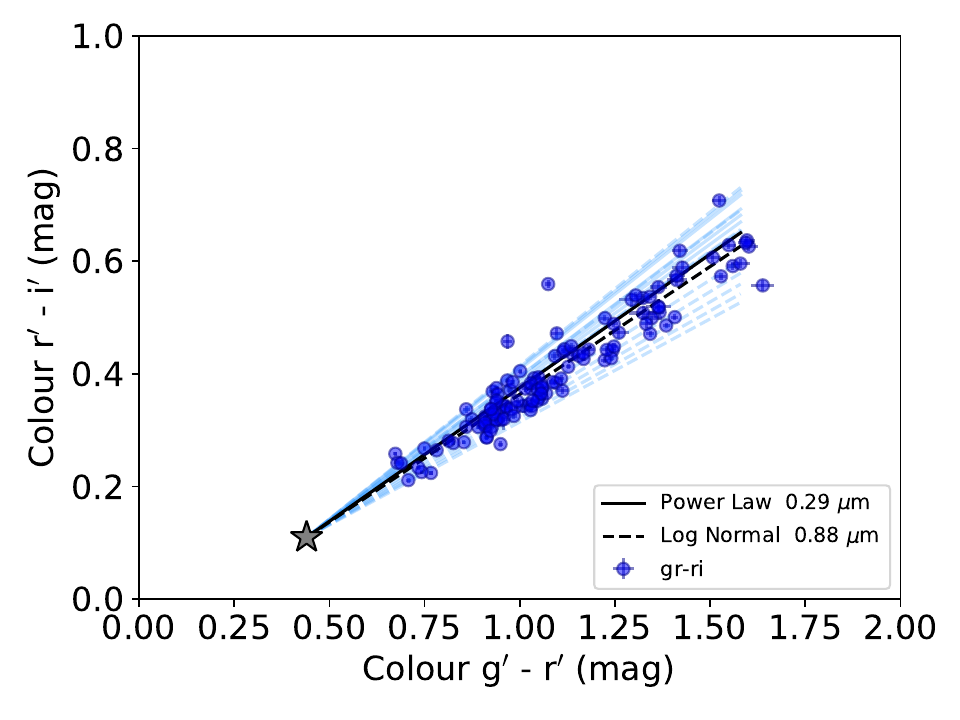}
    \includegraphics[width=0.48\textwidth]{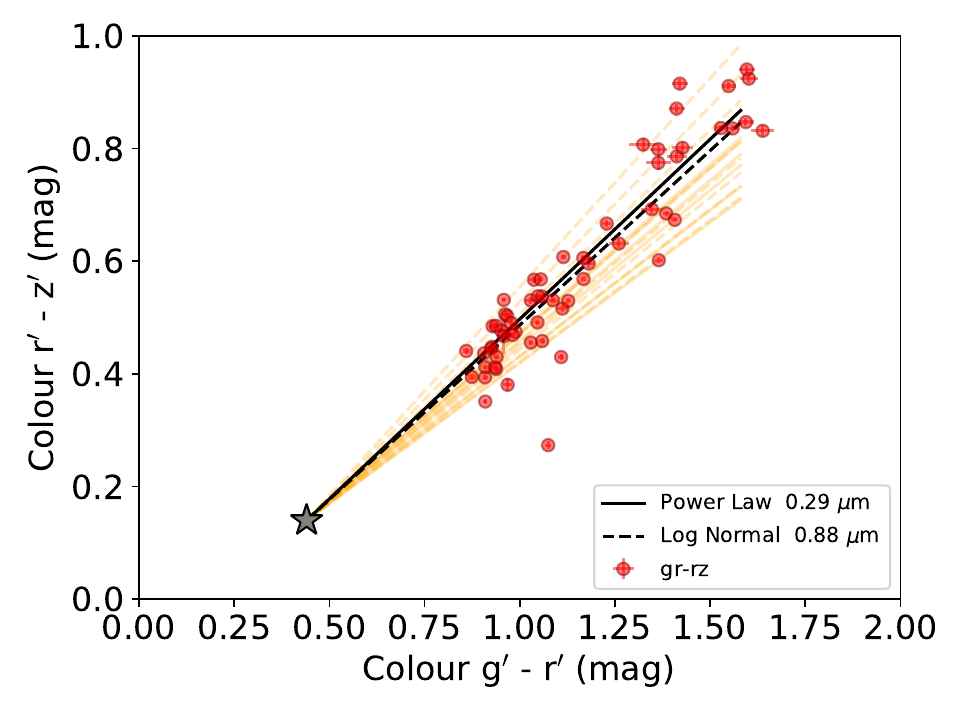}
    \caption{Colour-colour plots of ASASSN-21qj highlighting the fitting results for amorphous pyroxene (Mg$_{1.6}$Fe$_{0.4}$Si$_{2}$O$_{6}$); on the left is $g^{\prime}$-$r^{\prime}$ vs $r^{\prime}$-$i^{\prime}$ and on the right is $g^{\prime}$-$r^{\prime}$ vs $r^{\prime}$-$z^{\prime}$. In each panel the LCOGT photometry (colours) are denoted by circular data points, whilst the $\star$ symbol denotes the photospheric colours for a Sun-like star. We omit data points of the deepest occultations from this graph, focusing here on fitting the reddening due to the bulk of the dust. Reddening lines for power law (solid) and log normal (dashed) size distributions that fit the observations are shown as coloured lines. The best fit lines for each distribution are shown as black lines.}
    \label{fig:col_col_bf}
\end{figure*}

The data points associated with the two deepest occultation measurements $g^{\prime} \simeq 20~$mag, on MJDs 59476 and 59507, do not follow the same reddening vector as the majority of the data points. If we assume the grains in those two deep occultations have the same size distribution and composition as the remainder of the cloud, we can infer a minimum grain size of $\geq 1.2~\mu$m for the power law distribution, limited by the wavelength range from which the reddening vector is derived. This suggests some degree of grain size segregation within the occulting cloud. As the overall level of dimming decreases with time, this is consistent with the idea that smaller grains will spread more quickly from an impact or breakup, either through kinetic displacement or subsequent spreading by radiation forces, whereas larger grains may remain more tightly associated with their point of formation.

Based on the inferred power law size distribution, the dimming associated with the deepest occultations requires an intervening dust mass of $3.5\times10^{-12}~M_{\oplus}$. Assuming that the dust cloud lies on a circular orbit with a period of 30.9 days (semi-major axis of 0.19~au), based on the very deep dips early in the evolution of the event, the integrated occluding dust mass was at least 1.50~$~\pm~0.04 \times10^{-9}~M_{\oplus}$ (comparable to the total mass of Hale-Bopp) during the initial deep occultations. From the degree of dimming and the shape of the occultation, the dust cloud responsible must extend several stellar radii both vertically and horizontally, making this a lower limit to the total mass in the system.

\subsection{Infrared excess}

Since the dust cloud is predicted to lie close to the star, we searched archival infrared measurements for evidence of excess emission from the system, indicative of a substantial circumstellar disc. We find no evidence of excess at near- or mid-infrared wavelengths in archival observations spanning $>$10 years, including \textit{Spitzer}/IRAC measurements from GLIMPSE360 \citep{2008Whitney} and the AllWISE catalogue \citep{2010Wright}. The absence of significant infrared excess, along with its inferred large age, rules against ASASSN-21qj being surrounded by a misaligned dense, optically thick circumstellar disc \citep[e.g. ][]{2020Bredall,2020Andsell}, strengthening the case for an exocometary interpretation for the origin of its variability.

However, the NEOWISE observations reveal an intriguing brightening of the source several years prior to the onset of the occultation events we set out to monitor. This near-infrared brightening is first seen in measurements around MJD 58200, shortly after the optical brightening seen in the ASAS-SN photometry around MJD 58100. As of the most recent NEOWISE epoch around MJD 59900, the observed infrared excess has dissipated. The same material could therefore be responsible for both the reflected and emitted light seen in the time series observations such that whatever event caused the occultations we now observe was captured serendipitously. This near-infrared brightening is not seen in the NEOWISE measurements of any other exocomet host star, making the presence of a transient significant infrared excess unique to ASASSN-21qj. None of the other stars in close proximity to ASASSN-21qj (within 1$\arcmin$) exhibit a similar pattern in NEOWISE photometry, from which we rule out an instrumental origin. Assuming the excess emission to be intrinsic to ASASSN-21qj, we derive dust temperatures by fitting a blackbody curve to the averaged NEOWISE photometry at each epoch after subtraction of a scaled stellar photosphere model. The excess is consistent with temperatures in the range 1300 to 700~K, equivalent to blackbody distances of $\simeq$ 0.05 to 0.15~au from the star. Accounting for the absorption and emission properties of dust with the minimum size and composition derived previously (i.e. 0.29~$\mu$m amorphous pyroxene), we would expect distances of 0.12 to 0.24~au at the temperatures inferred from the infrared excess (1300$^{+2700}_{-1000}$ to 700$^{+100}_{-85}$~K).

An upper limit to the dust mass responsible for the unusual events reported here can then be inferred from the near-infrared excess. Radiative transfer modelling using {\sc Hyperion} \citep{2011Robitaille}, adopting the previously calculated power law size distribution, dust composition, and spatial distribution, requires $\simeq 2\times10^{-6}~M_{\oplus}$ of dust to replicate the excess.

Hot dust exozodiacal dust around main sequence stars has been identified in around 10 to 20\% of systems, correlated with the presence of an outer cool debris disc \citep{2006Absil, 2013Absil, 2014Ertel, 2016Ertel, 2021Absil}. The dust masses associated with these hot exozodis amounts to a few $10^{-9}~M_{\oplus}$ \citep{2017Kirchschlager}. Around very young stars, near-infrared excess and variability thought to be associated with terrestrial planet formation has been observed, with inferred dust masses around $10^{-4}~M_{\oplus}$ \citep{2015Meng,2019Su,2020Su,2022Su}. Although ASASSN-21qj is an old star, the mass of dust seen here is more comparable to the young planet-forming systems. 

\section{Summary and Conclusions}
\label{sec:con}

We have monitored the ongoing, deep occultation of the Sun-like star ASASSN-21qj in four filter bands at optical and near-infrared wavelengths for a duration of 282 days. Combining these new observations with ancillary data from publicly available monitoring programs, we characterised the onset and evolution of the ongoing dimming of this star. 

As the only star identified to date with a substantial, evolving infrared excess, a compelling case can be made for further characterisation of this system at near- and mid-infrared wavelengths with JWST. This will extend our knowledge of the planetesimal population behind the enigmatic little dippers through detailed determination of the dust size distribution and mineralogy.

We propose that sources which exhibit episodic dimming events, such as ASASSN 21qj, Boyajian's star \citep{2016Boyajian}, and TYC 8830-410-1 \citep{2021Melis}, but lack any signatures of youth, be called ``mature dippers''.

We summarize our main findings as follows:
\begin{enumerate}
    \itemsep0em
    \item[$\bullet$] The occultations of ASASSN-21qj exhibit quasiperiodic behaviour around 30.9 days, from which we infer the approximate distance of the occulting material to be 0.19$\pm$~0.04~au.
    \item[$\bullet$] The reddening of ASASSN-21qj is wavelength dependent, consistent with an occulting dust cloud composed of sub-micron refractory grains with a power law size distribution, minimum grain size of 0.29~$\mu$m, and an amorphous pyroxene composition.
    \item[$\bullet$] The occulting dust mass present around the star at the earliest, deepest part of the ongoing occultation is comparable to the total mass of a mid-sized Solar system comet (1.5$\times10^{-9}~M_{\oplus}$).
    \item[$\bullet$] Time evolution of the optical and near-infrared lightcurves of ASASSN-21qj are consistent with the collisional disruption of a massive planetesimal with subsequent spreading and clearing of the debris.
    \item[$\bullet$] The presence of a substantial infrared excess at near-infrared wavelengths is consistent with the inferred dust clump radial distance and suggests a much larger total dust mass in the system up to a few $\times10^{-6}~M_{\oplus}$, depending on the assumed spatial and size distribution.
\end{enumerate}

\begin{acknowledgments}

We thank the referee for their insightful and constructive comments to the paper.

This research has made use of the SIMBAD database, operated at CDS, Strasbourg, France. This research has also made use of NASA's Astrophysics Data System. 

This work makes use of observations from the Las Cumbres Observatory global telescope network.

This work has made use of data from the Asteroid Terrestrial-impact Last Alert System (ATLAS) project. The Asteroid Terrestrial-impact Last Alert System (ATLAS) project is primarily funded to search for near earth asteroids through NASA grants NN12AR55G, 80NSSC18K0284, and 80NSSC18K1575; byproducts of the NEO search include images and catalogs from the survey area. This work was partially funded by Kepler/K2 grant J1944/80NSSC19K0112 and HST GO-15889, and STFC grants ST/T000198/1 and ST/S006109/1. The ATLAS science products have been made possible through the contributions of the University of Hawaii Institute for Astronomy, the Queen’s University Belfast, the Space Telescope Science Institute, the South African Astronomical Observatory, and The Millennium Institute of Astrophysics (MAS), Chile.

JPM acknowledges research support by the Ministry of Science and Technology of Taiwan under grant MOST109-2112-M-001-036-MY3. CdB acknowledges support by Mexican CONAHCYT research grant FOP16-2021-01-320608. AR has been supported by the UK Science and Technology research Council (STFC) via the consolidated grant ST/S000623/1 and by the European Union’s Horizon 2020 research and innovation programme under the Marie Sklodowska-Curie grant agreement No. 823823 (RISE DUSTBUSTERS project). This work was also partly supported by the Spanish program Unidad de Excelencia Mar\'ia de Maeztu CEX2020-001058-M, financed by MCIN/AEI/10.13039/501100011033. SZ acknowledges support from the European Space Agency (ESA) as an ESA Research Fellow.

\end{acknowledgments}

%

\facilities{ASAS-SN, \textit{Gaia}, LCOGT, \textit{NEOWISE}}


\software{Astropy \citep{2013Astropy,2018Astropy},  Matplotlib \citep{2007Hunter}, NumPy \citep{2020Harris}, Photutils \citep{2022Bradley}, and SciPy \citep{2020Virtanen}, Optool \citep{2021Dominik}
          }






\bibliography{asassn21qj_refs}{}
\bibliographystyle{aasjournal}



\end{document}